\documentclass[%
superscriptaddress,
showpacs,
amsmath, amssymb,
aps,
twocolumn,
floatfix,
]{revtex4}

\usepackage{verbatim}
\usepackage{graphicx}
\usepackage{dcolumn}
\usepackage{bm}
\usepackage{hyperref}
\usepackage{subfigure}
\usepackage{float}
\usepackage{longtable}
\usepackage[tracking=true,kerning=true,expansion=true,spacing=false,factor=1100,stretch=20,shrink=20]{microtype}
\SetTracking{encoding={*}, shape=sc}{50}
\newcolumntype{P}[1]{>{\centering\arraybackslash}p{#1}}

\newcommand{\invfb}{\rm{fb}^{-1}}

\begin{document}
\title{Search for Higgs-like particles produced in association with bottom quarks in proton-antiproton collisions}
\affiliation{Institute of Physics, Academia Sinica, Taipei, Taiwan 11529, Republic of China}
\affiliation{Argonne National Laboratory, Argonne, Illinois 60439, USA}
\affiliation{University of Athens, 157 71 Athens, Greece}
\affiliation{Institut de Fisica d'Altes Energies, ICREA, Universitat Autonoma de Barcelona, E-08193, Bellaterra (Barcelona), Spain}
\affiliation{Baylor University, Waco, Texas 76798, USA}
\affiliation{Istituto Nazionale di Fisica Nucleare Bologna, \ensuremath{^{kk}}University of Bologna, I-40127 Bologna, Italy}
\affiliation{University of California, Davis, Davis, California 95616, USA}
\affiliation{University of California, Los Angeles, Los Angeles, California 90024, USA}
\affiliation{Instituto de Fisica de Cantabria, CSIC-University of Cantabria, 39005 Santander, Spain}
\affiliation{Carnegie Mellon University, Pittsburgh, Pennsylvania 15213, USA}
\affiliation{Enrico Fermi Institute, University of Chicago, Chicago, Illinois 60637, USA}
\affiliation{Comenius University, 842 48 Bratislava, Slovakia; Institute of Experimental Physics, 040 01 Kosice, Slovakia}
\affiliation{Joint Institute for Nuclear Research, RU-141980 Dubna, Russia}
\affiliation{Duke University, Durham, North Carolina 27708, USA}
\affiliation{Fermi National Accelerator Laboratory, Batavia, Illinois 60510, USA}
\affiliation{University of Florida, Gainesville, Florida 32611, USA}
\affiliation{Laboratori Nazionali di Frascati, Istituto Nazionale di Fisica Nucleare, I-00044 Frascati, Italy}
\affiliation{University of Geneva, CH-1211 Geneva 4, Switzerland}
\affiliation{Glasgow University, Glasgow G12 8QQ, United Kingdom}
\affiliation{Harvard University, Cambridge, Massachusetts 02138, USA}
\affiliation{Division of High Energy Physics, Department of Physics, University of Helsinki, FIN-00014, Helsinki, Finland; Helsinki Institute of Physics, FIN-00014, Helsinki, Finland}
\affiliation{University of Illinois, Urbana, Illinois 61801, USA}
\affiliation{The Johns Hopkins University, Baltimore, Maryland 21218, USA}
\affiliation{Institut f\"{u}r Experimentelle Kernphysik, Karlsruhe Institute of Technology, D-76131 Karlsruhe, Germany}
\affiliation{Center for High Energy Physics: Kyungpook National University, Daegu 702-701, Korea; Seoul National University, Seoul 151-742, Korea; Sungkyunkwan University, Suwon 440-746, Korea; Korea Institute of Science and Technology Information, Daejeon 305-806, Korea; Chonnam National University, Gwangju 500-757, Korea; Chonbuk National University, Jeonju 561-756, Korea; Ewha Womans University, Seoul, 120-750, Korea}
\affiliation{Ernest Orlando Lawrence Berkeley National Laboratory, Berkeley, California 94720, USA}
\affiliation{University of Liverpool, Liverpool L69 7ZE, United Kingdom}
\affiliation{University College London, London WC1E 6BT, United Kingdom}
\affiliation{Centro de Investigaciones Energeticas Medioambientales y Tecnologicas, E-28040 Madrid, Spain}
\affiliation{Massachusetts Institute of Technology, Cambridge, Massachusetts 02139, USA}
\affiliation{University of Michigan, Ann Arbor, Michigan 48109, USA}
\affiliation{Michigan State University, East Lansing, Michigan 48824, USA}
\affiliation{Institution for Theoretical and Experimental Physics, ITEP, Moscow 117259, Russia}
\affiliation{University of New Mexico, Albuquerque, New Mexico 87131, USA}
\affiliation{The Ohio State University, Columbus, Ohio 43210, USA}
\affiliation{Okayama University, Okayama 700-8530, Japan}
\affiliation{Osaka City University, Osaka 558-8585, Japan}
\affiliation{University of Oxford, Oxford OX1 3RH, United Kingdom}
\affiliation{Istituto Nazionale di Fisica Nucleare, Sezione di Padova, \ensuremath{^{ll}}University of Padova, I-35131 Padova, Italy}
\affiliation{University of Pennsylvania, Philadelphia, Pennsylvania 19104, USA}
\affiliation{Istituto Nazionale di Fisica Nucleare Pisa, \ensuremath{^{mm}}University of Pisa, \ensuremath{^{nn}}University of Siena, \ensuremath{^{oo}}Scuola Normale Superiore, I-56127 Pisa, Italy, \ensuremath{^{pp}}INFN Pavia, I-27100 Pavia, Italy, \ensuremath{^{qq}}University of Pavia, I-27100 Pavia, Italy}
\affiliation{University of Pittsburgh, Pittsburgh, Pennsylvania 15260, USA}
\affiliation{Purdue University, West Lafayette, Indiana 47907, USA}
\affiliation{University of Rochester, Rochester, New York 14627, USA}
\affiliation{The Rockefeller University, New York, New York 10065, USA}
\affiliation{Istituto Nazionale di Fisica Nucleare, Sezione di Roma 1, \ensuremath{^{rr}}Sapienza Universit\`{a} di Roma, I-00185 Roma, Italy}
\affiliation{Mitchell Institute for Fundamental Physics and Astronomy, Texas A\&M University, College Station, Texas 77843, USA}
\affiliation{Istituto Nazionale di Fisica Nucleare Trieste, \ensuremath{^{ss}}Gruppo Collegato di Udine, \ensuremath{^{tt}}University of Udine, I-33100 Udine, Italy, \ensuremath{^{uu}}University of Trieste, I-34127 Trieste, Italy}
\affiliation{University of Tsukuba, Tsukuba, Ibaraki 305, Japan}
\affiliation{Tufts University, Medford, Massachusetts 02155, USA}
\affiliation{Waseda University, Tokyo 169, Japan}
\affiliation{Wayne State University, Detroit, Michigan 48201, USA}
\affiliation{University of Wisconsin-Madison, Madison, Wisconsin 53706, USA}
\affiliation{Yale University, New Haven, Connecticut 06520, USA}

\author{T.~Aaltonen}
\affiliation{Division of High Energy Physics, Department of Physics, University of Helsinki, FIN-00014, Helsinki, Finland; Helsinki Institute of Physics, FIN-00014, Helsinki, Finland}
\author{S.~Amerio\ensuremath{^{ll}}}
\affiliation{Istituto Nazionale di Fisica Nucleare, Sezione di Padova, \ensuremath{^{ll}}University of Padova, I-35131 Padova, Italy}
\author{D.~Amidei}
\affiliation{University of Michigan, Ann Arbor, Michigan 48109, USA}
\author{A.~Anastassov\ensuremath{^{w}}}
\affiliation{Fermi National Accelerator Laboratory, Batavia, Illinois 60510, USA}
\author{A.~Annovi}
\affiliation{Laboratori Nazionali di Frascati, Istituto Nazionale di Fisica Nucleare, I-00044 Frascati, Italy}
\author{J.~Antos}
\affiliation{Comenius University, 842 48 Bratislava, Slovakia; Institute of Experimental Physics, 040 01 Kosice, Slovakia}
\author{G.~Apollinari}
\affiliation{Fermi National Accelerator Laboratory, Batavia, Illinois 60510, USA}
\author{J.A.~Appel}
\affiliation{Fermi National Accelerator Laboratory, Batavia, Illinois 60510, USA}
\author{T.~Arisawa}
\affiliation{Waseda University, Tokyo 169, Japan}
\author{A.~Artikov}
\affiliation{Joint Institute for Nuclear Research, RU-141980 Dubna, Russia}
% ...

\collaboration{CDF Collaboration}
\altaffiliation[With visitors from]{
\ensuremath{^{a}}University of British Columbia, Vancouver, BC V6T 1Z1, Canada,
\ensuremath{^{b}}Istituto Nazionale di Fisica Nucleare, Sezione di Cagliari, 09042 Monserrato (Cagliari), Italy,
\ensuremath{^{c}}University of California Irvine, Irvine, CA 92697, USA,
\ensuremath{^{d}}Institute of Physics, Academy of Sciences of the Czech Republic, 182~21, Czech Republic,
\ensuremath{^{e}}CERN, CH-1211 Geneva, Switzerland,
\ensuremath{^{f}}Cornell University, Ithaca, NY 14853, USA,
\ensuremath{^{g}}University of Cyprus, Nicosia CY-1678, Cyprus,
\ensuremath{^{h}}Office of Science, U.S. Department of Energy, Washington, DC 20585, USA,
\ensuremath{^{i}}University College Dublin, Dublin 4, Ireland,
\ensuremath{^{j}}ETH, 8092 Z\"{u}rich, Switzerland,
\ensuremath{^{k}}University of Fukui, Fukui City, Fukui Prefecture, Japan 910-0017,
\ensuremath{^{l}}Universidad Iberoamericana, Lomas de Santa Fe, M\'{e}xico, C.P. 01219, Distrito Federal,
\ensuremath{^{m}}University of Iowa, Iowa City, IA 52242, USA,
\ensuremath{^{n}}Kinki University, Higashi-Osaka City, Japan 577-8502,
\ensuremath{^{o}}Kansas State University, Manhattan, KS 66506, USA,
\ensuremath{^{p}}Brookhaven National Laboratory, Upton, NY 11973, USA,
\ensuremath{^{q}}Istituto Nazionale di Fisica Nucleare, Sezione di Lecce, Via Arnesano, I-73100 Lecce, Italy,
\ensuremath{^{r}}Queen Mary, University of London, London, E1 4NS, United Kingdom,
\ensuremath{^{s}}University of Melbourne, Victoria 3010, Australia,
\ensuremath{^{t}}Muons, Inc., Batavia, IL 60510, USA,
\ensuremath{^{u}}Nagasaki Institute of Applied Science, Nagasaki 851-0193, Japan,
\ensuremath{^{v}}National Research Nuclear University, Moscow 115409, Russia,
\ensuremath{^{w}}Northwestern University, Evanston, IL 60208, USA,
\ensuremath{^{x}}University of Notre Dame, Notre Dame, IN 46556, USA,
\ensuremath{^{y}}Universidad de Oviedo, E-33007 Oviedo, Spain,
\ensuremath{^{z}}CNRS-IN2P3, Paris, F-75205 France,
\ensuremath{^{aa}}Universidad Tecnica Federico Santa Maria, 110v Valparaiso, Chile,
\ensuremath{^{bb}}Sejong University, Seoul 143-747, Korea,
\ensuremath{^{cc}}The University of Jordan, Amman 11942, Jordan,
\ensuremath{^{dd}}Universite catholique de Louvain, 1348 Louvain-La-Neuve, Belgium,
\ensuremath{^{ee}}University of Z\"{u}rich, 8006 Z\"{u}rich, Switzerland,
\ensuremath{^{ff}}Massachusetts General Hospital, Boston, MA 02114 USA,
\ensuremath{^{gg}}Harvard Medical School, Boston, MA 02114 USA,
\ensuremath{^{hh}}Hampton University, Hampton, VA 23668, USA,
\ensuremath{^{ii}}Los Alamos National Laboratory, Los Alamos, NM 87544, USA,
\ensuremath{^{jj}}Universit\`{a} degli Studi di Napoli Federico II, I-80138 Napoli, Italy
}
\noaffiliation

\begin{abstract}
We report on a search for a spin-zero non-standard-model particle in proton-antiproton collisions collected by the Collider Detector at Fermilab at a center-of-mass-energy of 1.96 TeV. 
This particle, the $\phi$ boson, is expected to decay into a bottom-antibottom quark pair and to be produced in association with at least one bottom quark. The data sample consists of events with three jets identified as 
initiated by bottom quarks and corresponds to $5.4~\invfb$ of integrated luminosity. In each event, the invariant mass of 
the two most energetic jets is studied by looking for deviations from the multijet background, which is modeled using data. No evidence is found for such particle. Exclusion upper limits ranging from 20 to 2\,pb are set 
for the product of production cross sections times branching fraction for hypothetical $\phi$ boson with mass between 100 and 300 GeV/$c^2$. These are the most stringent constraints to date. 
\end{abstract}

\pacs{PACS}

\maketitle

\section{INTRODUCTION}

The discovery of a Higgs boson~\cite{CMS_H,ATLAS_H} completes the standard model (SM), but does not exclude the existence of yet-unknown particles that could provide direct indication of 
non-SM physics. Many extensions of the SM, for instance, predict particles decaying into quark pairs. 
Non-SM spin-0 resonances with SM Yukawa-like~\cite{couplings} couplings would decay predominantly to heavy quarks and, if their masses do not exceed twice the top-quark mass, mostly to bottom-antibottom ($b\bar{b}$) quark pairs. 
Such particles are foreseen, for example, in minimal supersymmetric extensions of the SM 
(MSSM)~\cite{MSSM}, where two scalar Higgs doublets exist, leading to five physical Higgs bosons, of which three are electrically neutral and collectively denoted as $\phi$. The $\phi$ boson particles would be 
produced preferably in association with a $b$ quark. The decay into $b\bar{b}$ pairs is expected to have a branching fraction of about $90\%$ in this model~\cite{HXS}. 
While the production cross section for SM Higgs bosons through vector-boson fusion in proton-antiproton ($p\bar{p}$) collisions at 1.96 TeV is $0.07\pm0.01$\,pb~\cite{PDG}, the cross section for the $\phi b$ process is calculated to be ${\cal{O}} (1)$\,pb~\cite{couplings}. In addition, scalar neutral particles with large couplings to 
$b$ quark are also predicted as mediators in dark-matter models~\cite{DM1,DM2}. Even for 
resonances with nonenhanced couplings to $b$ quarks, the sensitivity of searches with $b$ quarks in the final state is competitive, due to the distinctive final-state features that allow background reduction. 

The analysis described in this paper searches for massive particles decaying into $b\bar{b}$ pairs and produced in association with one or more $b$ quarks. The signal is searched for in final states with at least three 
$b$ quarks, where the requirement of the third $b$ quark is used to further suppress the multijet background, thus increasing the signal sensitivity. The requirement of a fourth $b$ quark is not considered, as its 
kinematic distributions fall outside the available acceptance resulting in lower signal efficiency.

Searches for such a process have been performed by the CDF~\cite{Tom_Wright} and the D0~\cite{D0} experiments at the Tevatron $p\bar{p}$ collider, as well as by the CMS experiment in $pp$ collisions at 
the Large Hadron Collider (LHC)~\cite{CMS_2016}.
The combined CDF and D0 result showed an excess of events of more than two standard deviations ($\sigma$) over the SM background prediction, compatible with the signal of a $100-150$ GeV/$c^2$ $\phi$ boson particle~\cite{PhibbTev}. 
The CMS collaboration has set exclusion limits for such particles as functions of the MSSM parameters. But, because of the higher collision energy, which leads to a larger multijet production rate, searches 
for a particle with mass smaller than 200$\,$GeV/$c^2$ at the LHC are limited by the difficulties in selecting online low-energy jets. This analysis investigates the reported $2\sigma$ deviation using completely independent data with the 
same $p\bar{p}$ initial state in the low-mass range of 100 to 300~GeV/$c^2$. 

The analysis presented in this paper is based on data from $p\bar{p}$ collisions at 1.96 TeV center-of-mass energy collected by the CDF II detector and corresponding to 5.4~$\invfb$ of integrated luminosity. The 
sample corresponds to the data collected after Spring 2008, when an ad-hoc online selection, which requires at least one jet identified as being initiated by a $b$ quark ($b$-jet) through a 
secondary-vertex algorithm~\cite{CDF_b-trigger}, was implemented. The offline analysis requires at least three $b$-jets. The relatively long $b$-quark lifetime provides distinctive features against backgrounds, strongly enhancing the sensitivity 
of the search.

The paper is organized as follows. In Sec.~\ref{sec:II}, the CDF II detector and the online data selection system are briefly described, while the data selection and the signal simulation are outlined in Sec.~\ref{sec:III}. 
Sec.~\ref{sec:IV} presents the data-driven background model. In Sec.~\ref{sec:V}, the fits to the data assuming the background-only hypothesis are described. Systematic uncertainties are summarized in Sec.~\ref{sec:VI}. The search 
for a massive particle is presented in Sec.~\ref{sec:VII}, and the results are discussed in Sec.~\ref{sec:VIII}. Finally, the main conclusions are summarized in Sec.~\ref{sec:IX}.

\section{THE CDF II DETECTOR}\label{sec:II}

The CDF II detector was an azimuthally and forward-backward symmetric apparatus located around one of the $p\bar{p}$ collision points at the Fermilab Tevatron collider. A detailed description of its design 
and performance is in Refs.~\cite{Acosta:2004yw,SecVtx}. Cylindrical coordinates are used to describe the event kinematics, in which $\varphi$ is the azimuthal angle, $\theta$ is 
the polar angle with respect to the proton beam, $r$ is the distance from the nominal beam line, and positive $z$ corresponds to the proton-beam direction, with the origin at the center of the 
detector. Pseudorapidity is defined as $\eta =-\ln(\tan(\theta/2))$. The transverse momentum of a particle is  defined as $p_{\text{T}}=p\sin(\theta)$ and  the  transverse  energy  as $E_{\text{T}}=E\sin(\theta)$.

A superconducting solenoidal magnet provided a magnetic field of 1.4 T oriented along the beam direction. Tracking devices placed inside the magnet measured charged-particle trajectories~(tracks). In particular, precise track 
measurements near the interaction point were provided by silicon-strip tracking detectors~\cite{CDF_silicon} in the polar range $|\eta|<1.1$. A 3.1\,m long cylindrical drift chamber~\cite{COT} provided full coverage over 
the range $|\eta|<1$.  
  
Particle energies were measured by calorimeters surrounding the solenoid and covering the region $|\eta|<3.6$: segmented lead-scintillator electromagnetic~\cite{CEM} and iron-scintillator hadronic~\cite{CHA} modules. 
%

An online selection system (trigger)~\cite{trig1,trig2} reduced the rate of events to be permanently recorded from 1.7\,MHz to 150\,Hz. The trigger system was organized in a three-level architecture. 
The first level~(L1) was based on custom-designed hardware that exploited low-resolution muon, track, and calorimeter information to produce a decision. Events selected by L1 were analyzed by the level 2~(L2) system, a 
combination of hardware and commercial processors where a partial event reconstruction was performed. The level 3~(L3) consisted of a large array of processors where data were read out and accepted events were sent to mass 
storage.

\section{DATA SELECTION AND SIGNAL DESCRIPTION}\label{sec:III}

The data sample used in this measurement was collected with an ad-hoc trigger optimized for the selection of events with $b$-jets. The trigger selection reached high signal purity by performing online $b$-jet tagging: 
the secondary vertex (SV), corresponding to the position where the $b$ hadron decays, is inferred from clusters of tracks displaced from the primary $p\bar{p}$ interaction vertex. 

At L1, at least two central ($|\eta| < 1.5$) calorimetric energy depositions (towers), with $E_{\mathrm{T}} \geq 5$\,GeV and two tracks having $p_\mathrm{T} > 2$\,GeV$/c$ were required. At L2, jets with $E_{\mathrm{T}} > 15$\,GeV and 
$|\eta| < 1.0$ were reconstructed using a fixed-cone algorithm with a radius parameter, $R$, of 0.7~\cite{JetClu}. 
At least two tracks with signed impact parameter $d_0>90~\mu$m matched to one of the jets had to be identified. The signed impact parameter is defined as $d_0 =R_b \sin(\varphi_b -\varphi) \approx R_b(\varphi_b -\varphi)$, where $R_b$ and $\varphi_b$ 
are the $b$-hadron decay length and azimuthal angle, respectively. At this stage, the $b$-hadron decay length in the transverse plane was required to be greater than $0.1$~cm.
At L3, the L2 requirements were applied to the offline-quality variables. A more detailed description of the online selection algorithm is in Ref.~\cite{CDF_b-trigger}. 
This trigger replaced the lower-purity trigger used in the previous CDF $\phi b$ search~\cite{Tom_Wright} and was sufficiently selective to remain online even with instantaneous luminosities of up to 
$3.0\times10^{32}$ cm$^{-2}$s$^{-1}$.

The offline selection requires at least three jets with $E_\text{T} > 22$\,GeV and $|\eta| < 1$, with energies corrected to account for detector and physics effects, such as the presence of 
inactive material in the calorimeters and multiple $p\bar{p}$ interactions per beam crossing, according to the standard CDF procedures~\cite{CDF_en_corr}. Each of the three jets is required to 
be associated with a secondary vertex identified by the {\sc SECVTX} $b$-tagging algorithm~\cite{SecVtx}, which assigns to each jet a positive or negative tag. If the secondary vertex is reconstructed inside 
the jet cone, the jet has a positive tag. If the secondary vertex is found on the opposite side of the primary vertex with respect to the jet direction, the jet has a negative tag. While most of the jets initiated by $b$ quarks 
are positively tagged, negatively tagged jets are predominantly initiated by light-flavor quarks in which a false secondary vertex is reconstructed based on resolution tails of the tracks. 

The sample with three positively-tagged jets constitutes the signal sample, and is referred to as the triple-tagged sample. The sample where two jets have a positive tag and the third 
jet has a negative tag is referred to as control sample. A sample with at least three jets with $E_\text{T} > 22$\,GeV and $|\eta| < 1$, but with the requirement of just two positively-tagged jets, is used to model the backgrounds and is referred to as the double-tagged sample. 

The $p\bar{p} \to \phi b +X$ signal is simulated using the {\sc Pythia} 6.216~\cite{pythia} Monte Carlo simulation with the CTEQ5L~\cite{CTEQ5L} set of parton distribution functions (PDF), and passed through the detector and trigger 
simulation based on a {\sc GEANT3}~\cite{GEANT3} description. At tree level, the cross section for this signal is dominated by the process $gg \to b \bar{b} H$. The process 
$gg\rightarrow b\bar{b}H$ is employed to simulate the signal final state. The standard model Higgs boson, forced to decay into a $b\bar{b}$ quark pair and with modified mass, is used to mimic the narrow $\phi$ state. 
Samples are generated for a variety of $\phi$ masses with a lower threshold of $15$~GeV$/c$ on the bottom quark $p_{\text{T}}$. 
These simulated signals are used to evaluate the acceptance and efficiency for reconstructing a $\phi b$ signal as functions of the $\phi$ mass. The combined efficiency and acceptance for the event selection 
increases from 0.37\% to 0.87\% for $\phi$ boson masses from 100~GeV$/c^2$ to 250~GeV$/c^2$, respectively, and then decreases down to 0.80\% at 300~GeV$/c^2$. At very high masses the efficiency decreases because 
the $b$ quarks produced in association are more likely to fall outside the acceptance.

\section{BACKGROUND DESCRIPTION}\label{sec:IV}

The dominant background is the multijet production of heavy-flavor quarks, which is conventionally categorized into the following processes: flavor creation, flavor excitation, and gluon splitting~\cite{bbcomposition}. Events where 
two gluon-splitting processes occur, or a flavor excitation process is followed by a gluon-splitting process, can lead to final states with three or more heavy quarks. 

The low-energy quantum chromodynamics (QCD) calculations that would be needed for reliable rate predictions of these events are intractable, thus it is not possible to rely on direct theoretical predictions. Furthermore, the invariant mass of the two leading-$E_{\mathrm{T}}$ jets, $m_{12}$, is affected by biases introduced by 
the trigger and the displaced-vertex tagging requirements that would need to be modeled. Therefore, a data-driven approach is chosen to model the various background components. 
Small ($<1\%$) contributions from $Z$ bosons produced in association with $b$-jets followed by $Z \to b\bar{b}$ decay, and from $t\bar{t}$ pair production, are neglected. 

The previous CDF measurement~\cite{Tom_Wright} showed that the triple-tagged jets sample contains predominantly two jets initiated by real $b$ quarks. Furthermore, the contamination from light-quark-initiated jets in the 
double-tagged sample is negligible as shown in Ref.~\cite{Zbb}, where the same online selection is used. Hence, the double-tagged sample is used to determine the 
normalized multijet-backgroud distributions (templates) needed for the analysis of the triple-tagged sample. The events in the double-tagged sample, with an additional third untagged jet, are separated into two categories, 
$bbY$ and $Ybb$, where $Y$ can take values ``$B$'' for bottom quark, ``$C$'' for charm quark, 
and ``$Q$'' for light quark or gluon. The classification label depends on the $E_{\text{T}}$ rank of 
the untagged jet, which is represented by the upper-case letter $Y$, and no distinction is made between the two leading jets. The sample where the third leading jet and either one of the two leading jets is tagged is labeled $Ybb$, while $bbY$ indicates 
events with an untagged third jet.

Six background templates, $bbB$, $Bbb$, $Cbb$, $bbC$, $Qbb$, and $bbQ$, are constructed by weighting the events by the probability that the untagged jet of a given $E_{\text{T}}$ would be identified as a $b$-jet by the {\sc SECVTX}-tagging 
algorithm, under the condition that it was initiated by a $b$, $c$, or light quark. These probabilities, called tagging matrices, are constructed on a per-jet basis, assuming that they do not depend on the event topology, but only on jet 
kinematic properties. They have been studied using simulated samples of $b\bar{b}$, $c\bar{c}$, and light-quark samples generated with the full CDF II detector simulation.

The simulated $b\bar{b}$ sample includes contributions from flavor creation, flavor excitation, and gluon splitting, while the $c\bar{c}$ sample is generated assuming only flavor creation. 
Differences in response of the online and the offline $b$-tagging algorithms between jets in experimental and simulated data are corrected using scale factors evaluated on a dedicated data sample~\cite{Zbb}. The value of the 
trigger scale factor is $0.68 \pm 0.03$, and for the offline $b$-tagging is $0.86 \pm 0.05$.
The $b$-tagging data-to-simulation scale factors are determined as functions of the jet $E_{\mathrm{T}}$ and applied to each simulated jet.

To further discriminate the jet-flavor composition of  the triple-tagged sample, a second variable, $x_{\text{tags}}$, is introduced alongside $m_{12}$. The $x_{\text{tags}}$ variable is derived from 
$M_{\text{SV}}$, the invariant mass of all tracks, assumed to be charged pions, associated with the reconstruction of the secondary vertex. 
The $M_{\text{SV}}$ distribution is sensitive to the flavor of the parton initiating the jet. For jets initiated by $c$ quarks, the distribution peaks at lower values than the one from jets initiated by $b$ quarks.
For the jets initiated by light quarks or gluons, denoted as $q$, a secondary vertex can only be reconstructed due to track mismeasurements. In this case, the $M_{\text{SV}}$ distribution follows an exponential decrease.
Following Ref.~\cite{Tom_Wright}, the $x_{\text{tags}}$ variable is defined as
\begin{equation}
\resizebox{.9\columnwidth}{!}{$\displaystyle
x_{\text{tags}} = \left \{  
\begin{array}{ccc}
min(M_{\text{SV},3}/\text{GeV}/c^2, 3) & : & M_{\text{SV},1} + M_{\text{SV},2} < 2~\text{GeV/}c^2 \\
min(M_{\text{SV},3}/\text{GeV}/c^2, 3)+3 & : & 2 < M_{\text{SV},1} + M_{\text{SV},2} < 4~\text{GeV/}c^2\\
min(M_{\text{SV},3}/\text{GeV}/c^2, 3)+6 & : & M_{\text{SV},1} + M_{\text{SV},2} > 4~\text{GeV/}c^2,\\
\end{array}
\right.
$}
\end{equation}

where $M_{\text{SV},1,2,3}$ is the $M_{\text{SV}}$ of the first, second, and third leading jet, respectively.
The $x_{\text{tags}}$ variable helps to discriminate backgrounds with high $M_{\text{SV}}$ from backgrounds with low $M_{\text{SV}}$. In particular, the $M_{\text{SV},1} + M_{\text{SV},2}$ distribution is sensitive to the $Cbb$ and $Qbb$ 
contributions, while the $M_{\text{SV},3}$ distribution discriminates statistically between the $bbC$ and $bbQ$ cases.

To build the $x_{\text{tags}}$ variable for the background templates, the events of the double-tagged sample are weighted by taking into account the flavor of the simulated untagged jet.
Because no SV is associated with the untagged jet in double-tagged events, for the computation of $x_{\text{tags}}$, all possible $M_{\text{SV}}$ values to the jet are assigned, each properly weighted by the 
tagging matrices, which are also parametrized as functions of the $M_{\text{SV}}$ variable. 
By construction, each event has multiple entries in the background template, each with the same value of $m_{12}$ and different $x_{\text{tags}}$. Since the number of events used to build the 
templates is two orders of magnitude larger than the yield of the analysis sample, the correlated fluctuations introduced in the $x_{\text{tags}}$ templates with this construction are neglected.

The $bbC$ and $bbQ$ template distributions are too similar to be discriminated by the fit. Therefore, their average distribution, $bbX$, is used, reducing the number of the
background templates to five. The $bbX$ double-tagged sample contains $1.3\times 10^5$ events 
and the $Ybb$ double-tagged sample contains $1.4\times 10^5$ events.

\section{RESULTS UNDER THE BACKGROUND-ONLY HYPOTHESIS}\label{sec:V}

The two-dimensional distribution in the variables $m_{12}$ and $x_{\text{tags}}$ for the 5\,616 triple-tagged events is fitted under the hypothesis that no signal is present.
A binned maximum-likelihood fit is used, where the likelihood function is constructed using a joint two-dimensional probability density function of the two variables $m_{12}$ and $x_{\text{tags}}$. 
The entries in each bin follow a Poisson distribution, $\mu^{n_{ij}}_{ij} e^{-\mu_{ij}}/n_{ij}!$, with $n_{ij}$ being the number of observed events in 
the $i$th bin of $m_{12}$ and the $j$th bin of $x_{\text{tags}}$, where the expected yield 
$\mu_{ij}$ is given by
\begin{equation}
 \mu_{ij} =\sum_b N_b f_{b,ij}.
\end{equation}

The index $b$ runs over the five background templates, $bbB$, $Bbb$, $Cbb$, $Qbb$, and $bbX$. The parameters $f_{b,ij}$ are the fractions contributed by each background component to bin $(i,j)$. The value $N_b$ of each 
background yield, normalized to the total number of events, is determined by the fit. 

The control sample, which consists of the $2\,359$ events with two positive and one negative $b$-tagged jets, is used to validate the background templates for light-flavor quarks. This sample, which 
is expected to contain almost purely $Qbb$ and $bbQ$ events, is fitted using all the background templates. 
The results return only contributions of the $Qbb$ and $bbX$ components, with $1\,701\pm132$ and $658 \pm 184$ events, respectively, with fit quality of $\chi^2$/d.o.f. = 26/22. The yields for the other three components are 
consistent with 0.

\begin{figure}
\centering
 \subfigure
{\includegraphics[width=1\columnwidth]{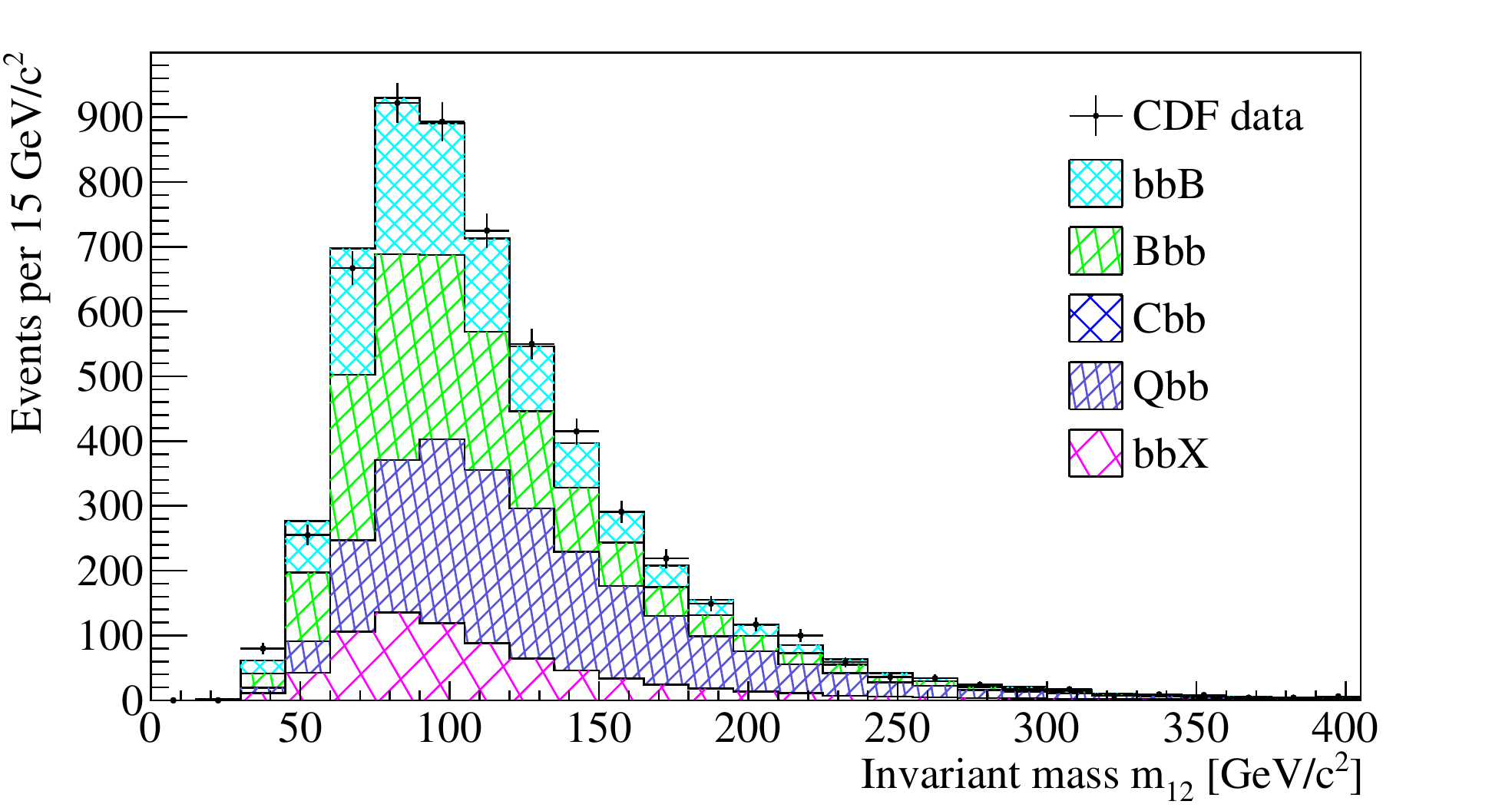}} 
\subfigure
{\includegraphics[width=1\columnwidth]{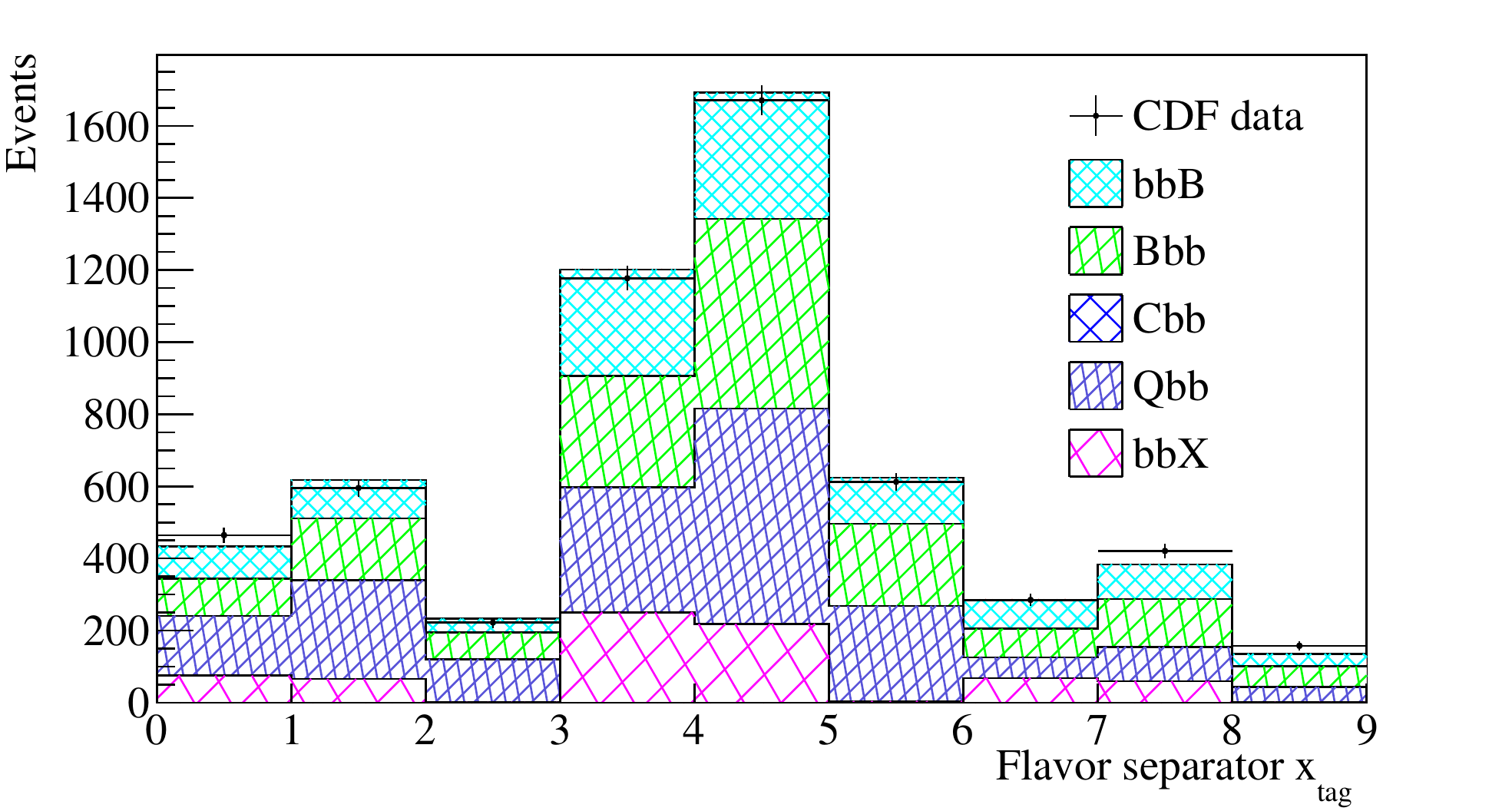}}
\caption{Triple-tagged events fit results projected into $m_{12}$ (top) and $x_{\text{tags}}$ (bottom), under the background-only hypothesis. The $Cbb$ component is found to be negligible.}
\label{fig:fit_bkg}
\end{figure}

The background templates are then used to fit the triple-tagged data sample. The result, projected onto the $m_{12}$ and $x_{\text{tags}}$ variables, is shown in Fig.~\ref{fig:fit_bkg}. No 
systematic uncertainties are included and the fit quality is $\chi^2/$d.o.f.$=17/22$.
Table~\ref{tab:fit_bkg} summarizes the fit results and compares them with an estimate based on the double-tagged sample. Studies using simulated samples in Ref.~\cite{Tom_Wright}, where the relevant analysis conditions 
mirror the present analysis, show that in events with at least two $b$-jets, about 2\% of the third jets are from $b$ quarks, about 4\% from $c$ quarks and the remaining from light quarks or
gluons, independently of the jet-energy ordering. The expected number of events for each 
background category in the triple-tagged sample is then estimated by multiplying the number of double-tagged events by these fractions. 

%
The expected numbers of $Qbb$ and $bbQ$ events of the $bbX$ template are extracted using the results of the fit to the negative-tagged control data sample. The results of the fit 
to the triple-tagged data sample assuming the background-only hypothesis are consistent with the predictions, with the exception of the $Cbb$ component, whose mass shape is too 
similar to the $bbB$ and $Bbb$ shapes to allow a significant separation by the fit. The large uncertainties in the $Bbb$ and $bbB$ fractions determined by the fit are due to their $-0.97$ correlation, which indicates 
that the fit is unable to distinguish between the two components. In the limit calculation described in Sec.~\ref{sec:VIII}, the correlation between background components is then taken into account. 

%

%

%
%
\begin{table*}[t]       
\centering
\caption{Event yields as determined by the fit to the triple-tagged sample in the background-only hypothesis, compared to the expectations based on extrapolating double-tagged yields using simulation-based 
 fractions (see text).}
  \begin{ruledtabular}
\begin{tabular}{c c c}

Background component & Best fit in the background-only hypothesis  & Expected yield from extrapolation \\ 

$bbB$ & $  1\,227 \pm  891$ & $950 \pm 48$\\
$Bbb$ & $  1\,672  \pm 738$ & $1\,280 \pm 64$\\
$Cbb$ & $ <90~(1\sigma)$ & $550 \pm 28$\\
$Qbb$ & $ 1\,964 \pm   169 $ & $1\,701 \pm 132$\\
$bbX$ & $ 742  \pm 293 $ & $658 \pm 184$\\

\end{tabular}                                                                 
\end{ruledtabular}
\label{tab:fit_bkg}                                                           
\end{table*}

\section{SEARCH FOR RESONANCES}\label{sec:VI}

\begin{figure}[!h]
\centering
 \subfigure
{\includegraphics[width=1\columnwidth]{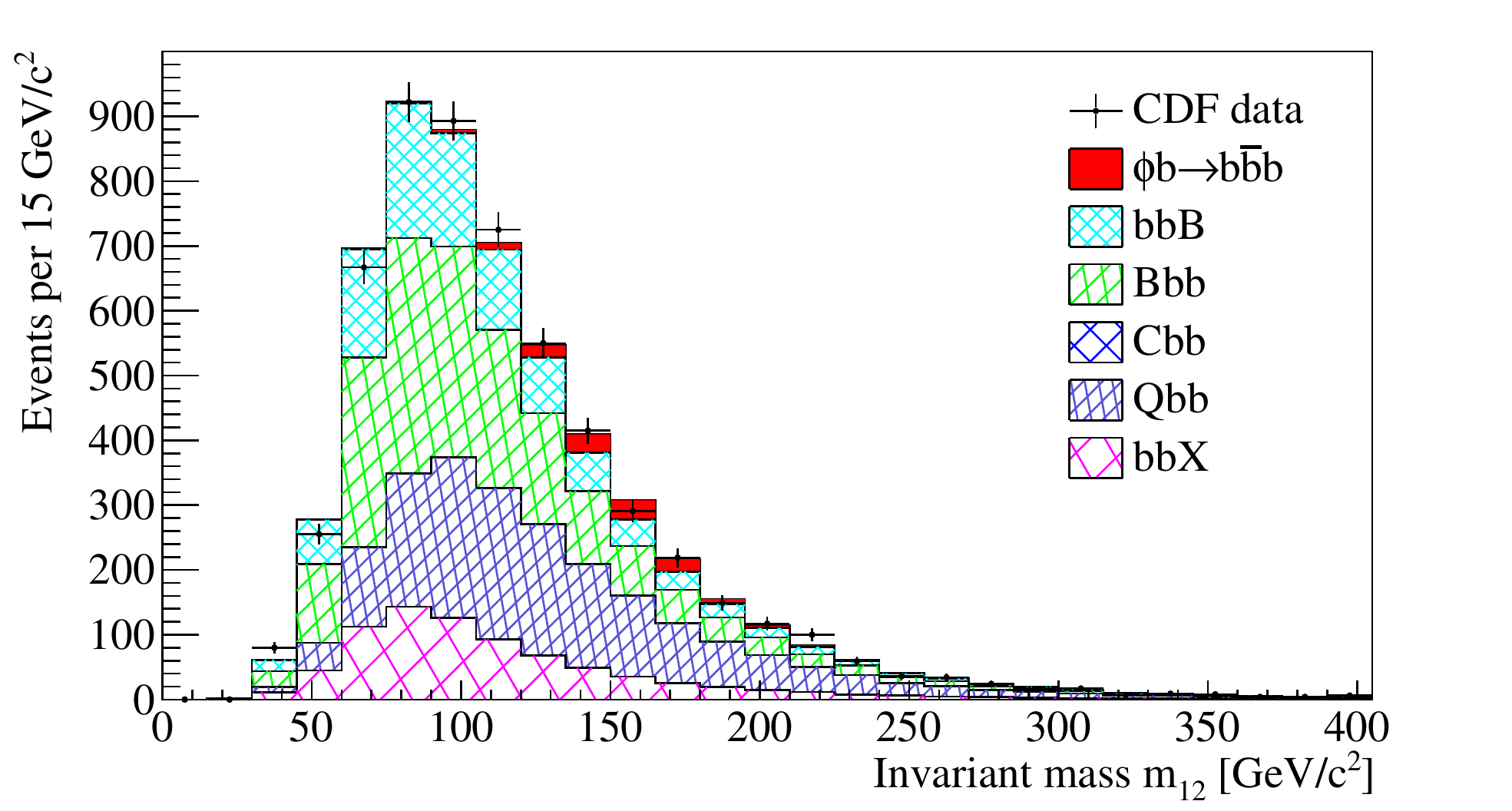}} 
\subfigure
{\includegraphics[width=1\columnwidth]{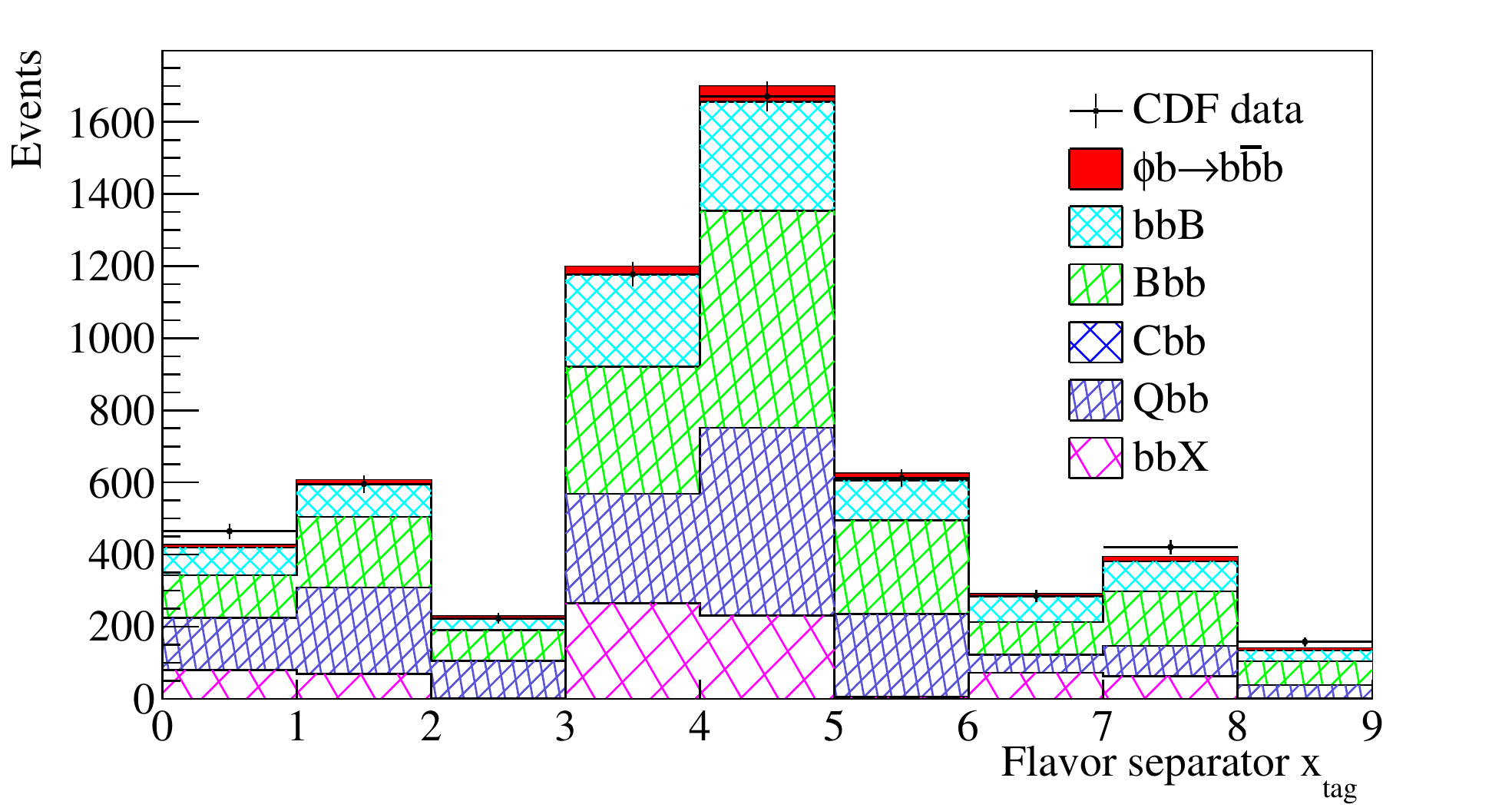}}
\caption{Result of the fit to the triple-tagged data projected into $m_{12}$ (top) and $x_{\text{tags}}$ (bottom). A signal component with a mass of the $\phi$ scalar of $160$\,GeV/$c^2$ is added to the background templates.}
\label{fig:fit_signal}

\end{figure}

A search for a Higgs-like particle $\phi$ is performed in the mass range of $100-300$\,GeV/$c^{2}$ by fitting the $m_{12}$ and the $x_{\text{tags}}$ distributions using the procedure described in the previous 
section and allowing for a signal component in the number of events in each bin $\nu_{ij}$
\begin{equation}
 \nu_{ij} =\sum_b N_b f_{b,ij} +  N_s f_{s,ij}.
\end{equation}

where $N_s$ is the total number of signal events, $f_{s,ij}$ represents the proportion of the signal template for each bin, and $N_b$ and $f_{b,ij}$ have the same meaning as in Sec~\ref{sec:V}.
The signal templates are obtained from the simulated signal samples with the requirement that three jets are $b$-tagged.
Figure~\ref{fig:fit_signal} shows the distributions of the leading dijets mass $m_{12}$ and the flavor separator $x_{tags}$, with results of the fit overlaid for a $\phi$ test mass of 160 GeV/$c^2$. In this case, 
the fit returns $130\pm70$ signal candidates, with a fit quality of $\chi^2/$d.o.f.$=16/21$. This would correspond to a cross section times branching fraction of about 7\,pb for the signal model, assuming a branching 
fraction of 90\% to $b\bar{b}$ quark pairs and a width of 36\,GeV/$c^2$. Only statistical uncertainties are considered here. 

Fits perfomed under various assumptions for the relative proportions of the $Cbb$, $Bbb$, and $bbB$ components yield consistent signal estimates, confirming that the similarity between background mass shapes prevents 
the fit from distinguishing precisely among various components but does not introduce signal biases.
\section{\label{sec:VII} SYSTEMATIC UNCERTAINTIES}

Systematic uncertainties affect both the signal and the background description. The uncertainties that impact the number of events of each component are classified as `rate' uncertainties, and the ones that come from the shape of the 
$m_{12}$ and $x_{\text{tags}}$ distributions are labeled as `shape' uncertainties. Table \ref{tab:sys} summarizes the systematic uncertainties considered. 

\begin{table}[h!]       
\centering
 \caption{Summary of the systematic uncertainties.}  
  \begin{ruledtabular}
\begin{tabular}{l c c c}          

Source &  Variation & Affects & Type \\ 

Luminosity & $5.9\%$ & Signal & Rate \\
Offline b-tag & 5$\%$ & Signal & Rate \\
Online b-tag & 4$\%$ & Signal & Rate\\
Jet energy scale & $4-7 \%$& Signal& Rate/shape \\
$x_{\text{tags}}$ & $3\%$ & Signal & Shape\\
PDFs & $2\%$ & Signal & Rate\\
Template stat.\ uncertainty & $-$  & Background & Shape \\

\end{tabular}                                                
\end{ruledtabular}
\label{tab:sys}              
\end{table}
The luminosity uncertainty follows Ref.~\cite{lumi}.
The online and offline $b$-tagging systematic uncertainties are taken from Ref.~\cite{Zbb}. The systematic uncertainty in the 
signal efficiency due to the CDF jet-energy correction is estimated by shifting the correction by 1\,$\sigma$ of its total uncertainty~\cite{JES}. In this way the acceptance and the shape of the 
signal are modified. The acceptance 
changes from $7\%$ to $4\%$ in the $100-300$~GeV/$c^2$ mass range of the $\phi$ particle.

The simulated signal samples are generated using the CTEQ5L set of PDFs. The uncertainty due to this choice is evaluated by generating simulated samples using the CTEQ6L~\cite{CTEQ6L} set and taking the difference in 
acceptance as uncertainty. The uncertainty due to the finite size of the background templates is taken into account assuming Poisson fluctuations in each bin.
The mass of the {\sc SECVTX} tags used to build the $x_{\text{tags}}$ variable, is varied by $\pm 3\%$ around the chosen values following Ref.~\cite{Tom_Wright}.

\section{LIMIT ON THE PRODUCTION CROSS SECTION}\label{sec:VIII}

The fitted signal yield in Sec.~\ref{sec:V} does not represent a clear evidence of a narrow states in the triple-tagged data set, whose composition is instead consistent with the sum of the background SM components. Exclusion upper limits at the $95\%$ confidence level (CL) on the production cross 
section times branching fraction are set as functions of the mass of the particle, by using a modified frequentist CL$_{\text{S}}$ method~\cite{CLs}. 
The limit calculation is based on the MCLIMIT package~\cite{MClimit}. Simulated experiments are generated based on the background modeling with the normalization taken from the third column of Table~\ref{tab:fit_bkg}, and 
on the various signal templates as functions of the $\phi$ mass. The fractions of the individual background normalizations and the signal yields are varied for each simulated experiment according to the systematic uncertainties in 
Table~\ref{tab:sys}. 

These simulated experiments are then fit under the background-only and the background-plus-signal hypotheses, with the $\phi$ mass varying between 100 and 300 GeV/$c^2$. 
The test statistic employed to calculate the limit is the difference in $\chi^2$ between the fits under the two hypotheses.  
The expected limit on the signal yield as a function of the $\phi$ mass is the median of the results in samples where no signal is present. The same procedure is repeated on data to determine the observed 
limit. 
The number of events is then translated into cross section times branching fraction, $\sigma(p\bar{p} \to \phi b){\cal{B}}(\phi \rightarrow b \bar{b})$, using the signal acceptance, the signal efficiency, the integrated 
luminosity, and the data-to-simulation scale factors for the online and offline $b$-tagging algorithm.

\begin{figure}[!h]
\centering

{\includegraphics[width=1\columnwidth]{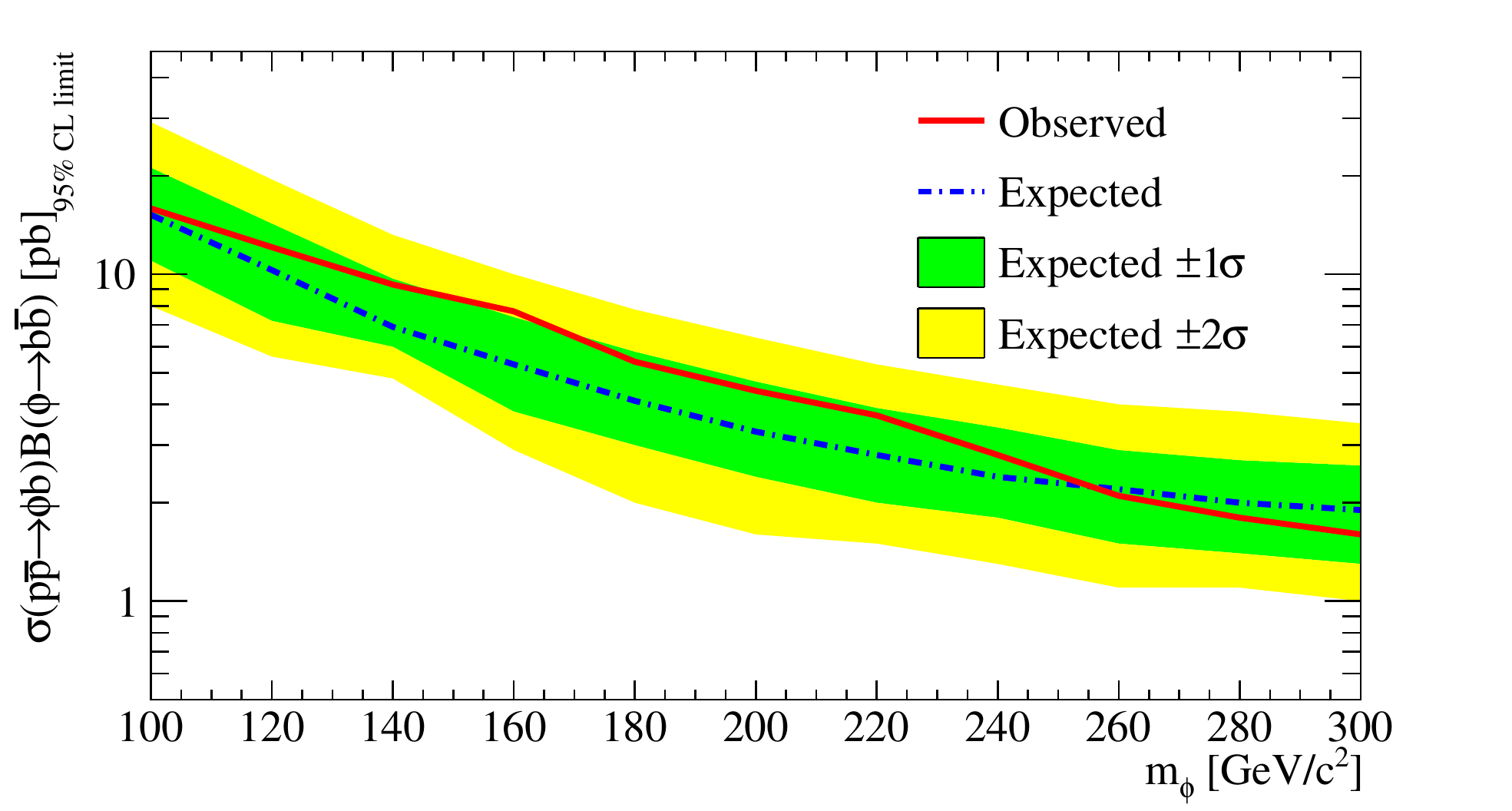}}
\caption{Observed and expected 95\% CL limits on $\sigma(p\bar{p} \to \phi b){\cal{B}}(\phi \rightarrow b \bar{b})$ as functions of $m_{\phi}$.}
\label{fig:limit}
\end{figure}

The observed 95\% CL limit, and the median expected limit under the background-only hypothesis, are summarized in Table~\ref{tab:limit} and shown in Fig.~\ref{fig:limit} 
with bands corresponding to fluctuations including 68.3\% ($1\sigma$) and 95.5\% ($2\sigma$) of the expected limits.

All observed limits are within the $1\sigma$ band of the expected limit, indicating the absence of any statistically significant excess of events. 

\begin{table}[!htb]   
\centering
\caption{Median expected and observed limits on $\sigma(p\bar{p} \to \phi b){\cal{B}}(\phi \rightarrow b \bar{b})$.}   
\begin{ruledtabular}
\begin{tabular}{c c c}

           & \multicolumn{2}{c}{95\% CL upper limit~[pb]} \\
$m_{\phi}~[\text{GeV}/c^2]$ & Expected  & Observed\\ 

100  & 15.2 & 15.9 \\
120  & 10.3 & 12.1 \\
140  & 6.9 & 9.3\\
160  & 5.3 & 7.7 \\
180  & 4.1 & 5.4\\
200  & 3.3 & 4.4\\
220  & 2.8 & 3.7 \\
240  & 2.4 &  2.8\\
260  & 2.2 & 2.1 \\
280  & 2.0 & 1.8 \\
300  & 1.9 & 1.6 \\

\end{tabular}
 \end{ruledtabular}
\label{tab:limit}  
\end{table}

\section{CONCLUSION}\label{sec:IX}

A search for a Higgs-like particle with 100-300 GeV/$c^2$ mass range decaying into a pair of $b$ quarks and produced in association with at least one additional $b$ quark in $p\bar{p}$ collisions is reported. 

No significant deviations from the SM expectations for background are observed. The sensitivity of this analysis is doubled with respect to the previous CDF result. 
For that analysis~\cite{Tom_Wright}, the most significant excess of events with respect to the expected background, was observed at $m_{\phi}= 150$ GeV/$c^2$ with a significance of 2.8\,$\sigma$. 
This excess, interpreted as associated to a narrow scalar particle, corresponded to a production cross section times branching fraction of about 15 pb. The result reported here excludes such a signal rate 
with 95\% confidence.

\section*{ACKNOWLEDGEMENTS}
This document was prepared by the CDF collaboration using the resources of the
Fermi National Accelerator Laboratory (Fermilab), a U.S.  Department of Energy,
Office of Science, HEP User Facility. Fermilab is managed by Fermi Research
Alliance, LLC (FRA), acting under Contract No.  DE-AC02-07CH11359.  We thank
the Fermilab staff and the technical staffs of the participating institutions
for their vital contributions. This work was supported by the U.S. Department
of Energy and National Science Foundation; the Italian Istituto Nazionale di
Fisica Nucleare; the Ministry of Education, Culture, Sports, Science and
Technology of Japan; the Natural Sciences and Engineering Research Council of
Canada; the National Science Council of the Republic of China; the Swiss
National Science Foundation; the A.P. Sloan Foundation; the Bundesministerium
f\"ur Bildung und Forschung, Germany; the National Research Foundation of Korea
(Grants No. 2018R1A6A1A06024970); the Science and Technology Facilities Council
and the Royal Society, United Kingdom; the Russian Foundation for Basic
Research; the Ministerio de Ciencia e Innovaci\'{o}n, and Programa
Consolider-Ingenio 2010, Spain; the Slovak R\&D Agency; the Academy of Finland;
and the Australian Research Council (ARC).


\begin{thebibliography}{usrt}
\bibitem{CMS_H} S. Chatrchyan {\it et al.} (CMS Collaboration), Phys. Lett. B {\bf 716}, 30 (2012).
\bibitem{ATLAS_H} G. F. Aad {\it et al.} (ATLAS Collaboration), Phys. Lett. B {\bf 716}, 1 (2012).
\bibitem{couplings} S. Dawson, C. B. Jackson, L. Reina, and D. Wackeroth, Mod. Phys. Lett. A {\bf 21}, 89 (2006).
\bibitem{MSSM} H. P. Nilles, Phys. Rept. {\bf 110}, 1 (1984).
\bibitem{HXS} D. de Florian {\it et al.},  arXiv:1610.07922.
\bibitem{PDG} C. Patrignani {\it et al.} (Particle Data Group), Chin. Phys. C {\bf 40}, 100001 (2016) and 2017 update.
\bibitem{DM1} E. Izaguirre, G. Krnjaic, and B. Shuve, Phys. Rev. D {\bf 90}, 055002 (2014).
\bibitem{DM2} A. Berlin, D. Hooper, and S. D. McDermott, Phys. Rev. D {\bf 89}, 115022 (2014).
\bibitem{Tom_Wright} T. Aaltonen {\it et al.} (CDF  Collaboration),  Phys.  Rev. D {\bf 85}, 032005 (2012).
\bibitem{D0} V. M. Abazov {\it et al.} (D0 Collaboration), Phys. Lett. B {\bf 698}, 97 (2011).
\bibitem{CMS_2016} V. Khachatryan {\it et al.} (CMS Collaboration), J. High Energy Phys. 11 (2015) 071.
\bibitem{PhibbTev} T. Aaltonen {\it et al.} (CDF and D0 Collaborations), Phys. Rev. D {\bf 86}, 091101 (2012).
\bibitem{CDF_b-trigger} S. Amerio, M. Casarsa, G. Cortiana, J. Donini, D. Lucchesi, and S. P. Griso, IEEE Trans. Nucl. Sci. {\bf 56}, 1690 (2009).
\bibitem{Acosta:2004yw} D. Acosta {\it et al.} (CDF Collaboration), Phys. Rev. D {\bf 71}, 032001 (2005).
\bibitem{SecVtx} D. Acosta {\it et al.} (CDF Collaboration), Phys. Rev. D {\bf 71}, 052003 (2005).
\bibitem{CDF_silicon} T. Aaltonen {\it et al.} (CDF Collaboration), Nucl. Instrum. Methods A {\bf 729}, 153 (2013).
\bibitem{COT} A. A. Affolder {\it et al.} (CDF Collaboration), Nucl. Instrum. Methods A {\bf 526}, 249 (2004).
\bibitem{CEM} L. Balka {\it et al.} (CDF Collaboration), Nucl. Instrum. Methods A {\bf 267}, 272 (1988).
\bibitem{CHA} S. Bertolucci {\it et al.} (CDF Collaboration), Nucl. Instrum. Methods A {\bf 267}, 301 (1988).
\bibitem{trig1} R. Downing {\it et  al.}, Nucl. Instrum. Methods Phys. Res., Sec. A {\bf 570}, 36 (2007).
\bibitem{trig2} E.J.  Thomson {\it et  al.},  IEEE  Trans.  Nucl.  Sci. {\bf 49},  1063(2003).
\bibitem{JetClu} F. Abe {\it et al.} (CDF Collaboration), Phys. Rev. D {\bf 45}, 1448 (1992). 
\bibitem{CDF_en_corr} A. Bhatti {\it et al.} (CDF Collaboration), Nucl. Instrum. Methods A {\bf 566}, 375 (2006).
\bibitem{pythia} T. Sj{\"o}strand, S. Mrenna,  and P. Skands, J. High Energy Phys. 05 (2006) 026.
\bibitem{CTEQ5L} H. L. Lai {\it et al.} (CTEQ Collaboration),  Eur. Phys. J. C {\bf 12}, 375 (2000).
\bibitem{GEANT3} R. Brun, R. Hagelberg, M. Hansroul, and J.C. Lassalle, CERN-DD-78-2 (1978).
\bibitem{bbcomposition} D. Acosta {\it et al.} (CDF Collaboration), Phys. Rev. D {\bf 71}, 092001 (2005).
\bibitem{Zbb} T. Aaltonen {\it et al.} (CDF Collaboration), Phys. Rev. D {\bf 98}, 072002 (2018).
\bibitem{lumi} D. Acosta {\it et al.}, Nucl. Instrum. Methods A {\bf 494}, 57 (2002).
\bibitem{CTEQ6L} J. Pumplin {\it et al.}, J. High Energy Phys. 07 (2002) 012.
\bibitem{JES} A. Bhatti {\it et al.}, Nucl. Instrum. Methods A {\bf 566}, 375 (2006).
\bibitem{CLs} A. L. Read, J. Phys. G {\bf 28}, 2693 (2002).
\bibitem{MClimit}  T. Junk, Nucl. Instrum. Methods A {\bf 434}, 435 (1999).
\end{thebibliography}

\end{document}